\documentclass[reprint]{revtex4-1}

\usepackage[pdftex]{graphicx}
\graphicspath{{./}}
\usepackage{amssymb,amsmath}

\newcommand{\Thomson}{\sigma_{\scriptscriptstyle\!T}}
\newcommand{\PL}{P_{\scriptscriptstyle\!L}}
\newcommand{\PC}{P_{\scriptscriptstyle\!C}}
\newcommand{\ZR}{Z_{\scriptscriptstyle\!R}}
\newcommand{\labar}{{\mkern0.75mu\mathchar '26\mkern -9.75mu\lambda}}

\begin{document}

\title{Radiation Damping by Thomson Scattering}
\author{Nickolai Muchnoi} \email{N.Yu.Muchnoi@inp.nsk.su}
\affiliation{Budker Institute of Nuclear Physics of Siberian Branch Russian Academy of Sciences}
\affiliation{Novosibirsk State University}

\begin{abstract}

Synchrotron radiation of relativistic electrons in storage rings naturally leads to the damping of betatron oscillations. 
Damping time and transverse beam emittance can be reduced by wigglers or undulators while the beam parameters are still well defined by the common radiation integrals, based on the properties of synchrotron radiation.
However the quantum excitation of betatron oscillations in principle can be considerably reduced if radiation occurs due to the Thomson scattering in the periodic electromagnetic field.
After a brief introduction we compare radiation properties for different cases and suggest the corresponding modification of the radiation integrals.
\end{abstract}

\maketitle 

\section{Introduction}

Radiation of relativistic electrons from the point of view of quantum electrodynamics occurs due to their scattering on the real or virtual photons.
After introducing the basic properties of the Compton effect we come to the equivalence of the Thomson scattering and the radiation of electrons in a weak harmonic undulator.
The scattering probability is defined as a product of the Thomson cross section and the density of photons along the electron propagation trajectory.
Consequently the changes in average energy, energy spread and the transverse emittance of the electron beam are evaluated as the product of the number of scattering acts and the average change of the corresponding parameter in a single scattering.
Within this approach we derive the radiation damping effect from undulator radiation and compare the results with those obtained by the application common radiation integrals~\cite{Chao:384825} for the case of a wiggler/undulator.
This comparison shows that the common radiation integrals only include the dispersion mechanism for excitation of betatron and synchrotron oscillations.
At present the radiation damping by wigglers is a well established technique~\cite{WIEDEMANN198824,PhysRevSTAB.15.042401} but also there were several suggestions to use a laser backscattering for the same purpose~\cite{PhysRevLett.78.4757,PhysRevLett.80.976,Lebedev:2004yz}.
The aim of the present study is to find the universal description of the harmonic damping devices for an arbitrary combination of the field strength and the field period (or wavelength).

\begin{figure}[h]
\centering
\includegraphics[width=0.9\linewidth]{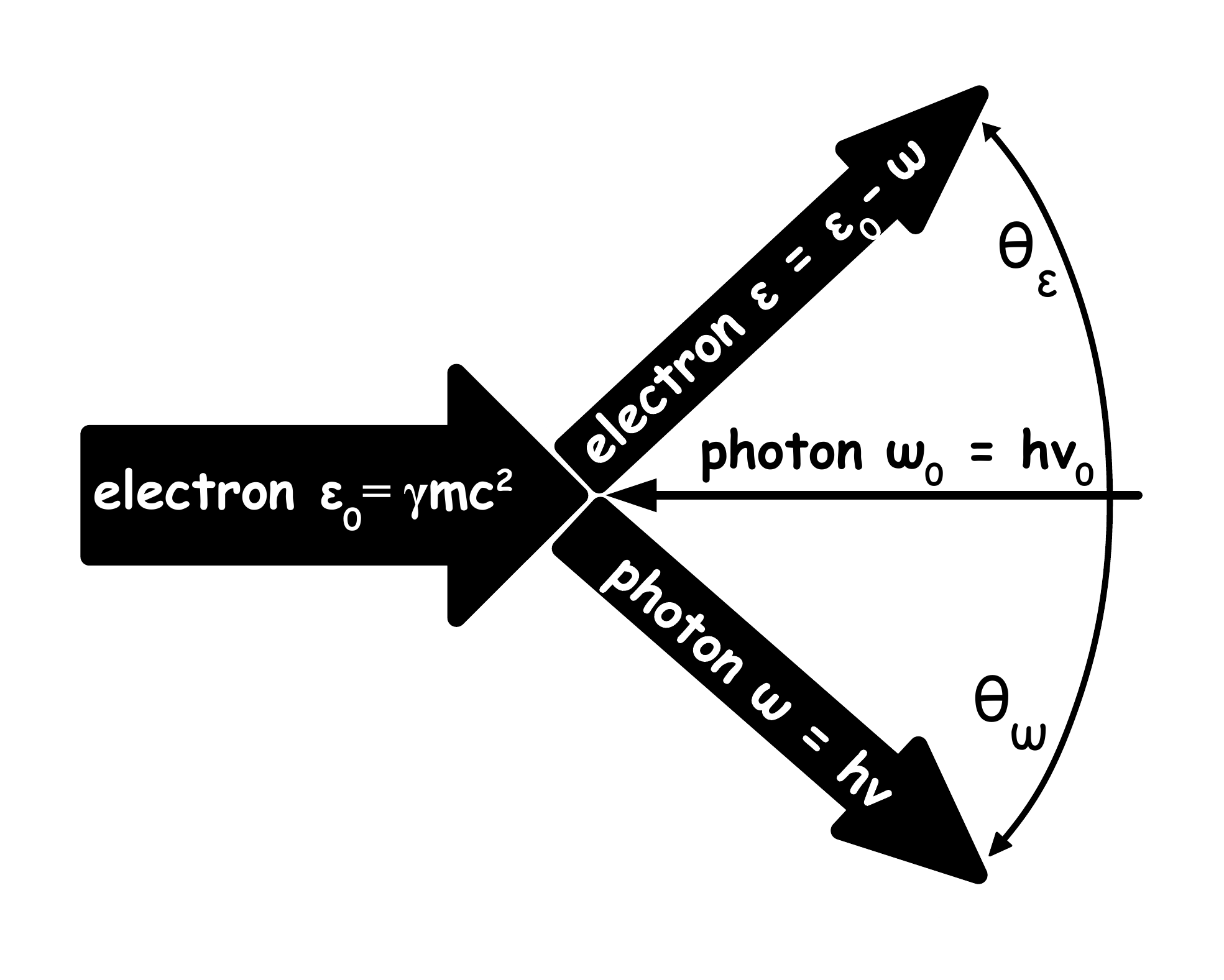}
\caption{Inverse Compton scattering: the thickness of the arrows qualitatively represents the energies of the particles.}
\label{fig:compton}
\end{figure}

Let us start by consideration of the scattering of an ultra-relativistic electron in the oncoming plane monochromatic electromagnetic wave.
In Fig.~\ref{fig:compton} there is an illustration for the process of inverse Compton scattering of a soft photon with energy $\omega_0$ on an ultra-relativistic electron with energy $\varepsilon_0=\gamma m c^2$. 
The electron transfers a part of its energy $\omega\gg\omega_0$ to the photon so that the energy conservation law looks like $\varepsilon_0 \simeq \varepsilon + \omega$.

We take from~\cite{berestetskii1982quantum} the cross section for scattering of linearly polarised photons on unpolarised electrons at rest, summed over the polarisation states of the final photons. 
For our case, after Lorentz transformations we have:
\begin{equation}
\frac{d\sigma}{du\,d\varphi}  =
\frac{r_e^2}{\kappa(1+u)^2}\!
\left(2\!+\!\frac{u^2}{1+u}\!+\!8\frac{u}{\kappa}\!\left[\frac{u}{\kappa}-1\right]\sin^2\!\varphi\!\right)\!,
\label{dsdudfi}
\end{equation}
where $\kappa=4\gamma\omega_0/mc^2$ is twice the ratio between the photon and electron energies in the electron rest frame and 
$u\in[0,\kappa]$ is the dimensionless parameter \cite{PhysRevLett.110.140402}:
\begin{equation}
 u  = \displaystyle\frac{\omega}{\varepsilon} 
     =\frac{\theta_\varepsilon}{\theta_\omega} 
     =\frac{\omega}{\varepsilon_0-\omega}
     =\frac{\varepsilon_0-\varepsilon}{\varepsilon}.
\label{u}
\end{equation}
$\varphi$ is the angle between the plane of polarisation and the scattering plane.
$\theta_\omega$ and $\theta_\varepsilon$ in Fig.~\ref{fig:compton} are defined as:
\begin{equation}
\theta_\omega(u)      = \frac{1}{\gamma}\sqrt{\frac{\kappa}{u}-1};\hspace{2em}
\theta_\varepsilon(u) = \frac{u}{\gamma}\sqrt{\frac{\kappa}{u}-1}\;.
\label{nwne}
\end{equation}
The energy of the scattered photon and its maximum possible value are:
\begin{equation}
\omega(u)    = \varepsilon_0\frac{u     }{1+u     };\hspace{2em}
\omega_{max} = \varepsilon_0\frac{\kappa}{1+\kappa}\;.
\label{wwmax}
\end{equation}
In the limiting case when $\kappa\ll1$ the approximations $\omega_{max} \simeq \varepsilon_0 \kappa$, $du \simeq d\omega/\varepsilon_0$ and $u/\kappa \simeq \omega/\omega_{max}$ allow to rewrite Eq.~(\ref{dsdudfi}) as:
\begin{equation}
\frac{d\sigma}{d\omega\,d\varphi}  =
\frac{2r_e^2}{\omega_{max}}\!
\left(1+4\frac{\omega}{\omega_{max}}\left[\frac{\omega}{\omega_{max}}-1\right]\sin^2\varphi\right).
\label{dsdwdfi}
\end{equation}
Integration of (\ref{dsdwdfi}) over $\varphi$ yields: 
\begin{equation}
\frac{d\sigma}{d\omega}  =
\frac{4\pi r_e^2}{\omega_{max}}\!
\left(1-2\frac{\omega}{\omega_{max}}+2\left[\frac{\omega}{\omega_{max}}\right]^2\right).
\label{dsdw}
\end{equation}
Integration of (\ref{dsdw}) in the range from 0 to $\omega_{max}$ gives the Thomson cross section $\Thomson=8 \pi r_e^2/3$.
Thus the case of scattering with $\kappa\ll1$ may be considered as Relativistic Thomson Scattering (RTS).
Further we will need averaged values of some RTS parameters, which can be easily obtained using the definitions (\ref{nwne}), (\ref{wwmax}) and the cross section expression (\ref{dsdudfi}).
The lowest order in the parameter $\kappa$ yields the following results:
\begin{eqnarray}
\left< \omega \right>     & = & 2\gamma^2\omega_0, \label{wave}\\
\left< \omega^2 \right>   & = & \frac{28}{5} (\gamma^2\omega_0)^2 = \frac{7}{5}\left< \omega \right>^2, \label{w2ave}\\
\left<\left(\theta_\varepsilon\cos\varphi\right)^2\right> & = & \;\;\frac{6}{5}\left(\frac{\lambda_c}{\lambda_0}\right)^2 = \frac{1}{2}\left<\theta_\varepsilon^2\right>,\label{txave}\\
\left<\left(\theta_\varepsilon\sin\varphi\right)^2\right> & = & \;\;\frac{4}{5}\left(\frac{\lambda_c}{\lambda_0}\right)^2 = \frac{1}{3}\left<\theta_\varepsilon^2\right>,\label{tyave}
\end{eqnarray}
where $\lambda_0$ is the wavelength of incident photon and $\lambda_c$ is the Compton wavelength of an electron.

\section{Relativistic Thomson Scattering}
\subsection{Laser Radiation}

Let us explore a case when an ultra-relativistic electron propagates along the axis of a 100\% linearly polarised gaussian CW laser beam in the direction, opposite to the light propagation.
If the electron is on-axis of the laser beam within $z\in[-a:a]$, the probability of scattering is the product of the Thomson cross section $\Thomson$ and the density of the photon target: 
\begin{equation}
W = \Thomson\frac{\PL \lambda_0}{\pi h c^2}\!\!\int\limits_{-a}^{a}\!\!\frac{dz}{\sigma(z)^2}
=\frac{\PL}{\PC} \frac{\arctan(a/\ZR)}{\pi/2},
\label{WCW}
\end{equation}
where
\begin{itemize}
\item $\PL$~[W] and $\lambda_0$~[m] are the laser power and the wavelength of its radiation. 
$\PL\lambda_0/hc^2$~[m$^{-1}$] is the longitudinal density of laser photons.
\item $\sigma(z) = \sigma_0\sqrt{1+(z/\ZR)^2}$ is the transverse RMS laser beam size with the waist size $\sigma_0$ at location $z=0$.
$\ZR = 4\pi\sigma_0^2/\lambda_0$ is the Rayleigh length.
\item $\PC = \hbar c^2/2\Thomson \simeq 0.7124\cdot10^{11}$~[W] is the critical laser power when the probability of scattering is close to 100\% (assuming $a\gg\ZR$).
\end{itemize}

According to Eq.~(\ref{WCW}) if $P_L=1$~[W], the scattering probability $W\lesssim1.4\cdot10^{-11}$ and it does not depend on laser radiation wavelength.
The RTS of intense laser pulse had been proposed in \cite{PhysRevLett.80.976} for an electron beam cooling in a low-energy storage ring.
In \cite{DeJongh2001585} this approach was suggested for cooling of TeV muons.
For laser cooling an extremely high laser power is required to produce a sufficiently strong field in the scattering area.
The power of a gaussian laser beam is coupled with the on-axis magnetic induction amplitude $B_0$~[T] as:
\begin{equation}
\PL  =  \frac{\pi c}{\mu_0} \sigma(z)^2 B_0(z)^2,
\label{PLB0}
\end{equation}
where $\mu_0=4\pi\cdot10^{-7}$~[N A$^{-2}$].

\subsection{Undulator}

Now consider a plane undulator with a period $\lambda_u$ $(\labar_u=\lambda_u/2\pi)$ and on-axis magnetic induction $B_y(z)=B_u\sin(z/\labar_u)$.
The dimensionless undulator parameter is:
\begin{equation}
K  =  \frac{e B_u \labar_u}{m c} = \frac{B_u}{B_c}\frac{\labar_u}{\labar_c},
\label{K}
\end{equation}
where $B_c \labar_c = mc/e \simeq 1.7\cdot10^{-3}$~[T~m] is the product of the Schwinger field strength $B_c = 4.414\cdot10^{9}$~[T] and the Compton wavelength of the electron.
The maximum photon energy of the basic harmonic of undulator radiation is:
\begin{equation}
\omega_{max} = \frac{2 \gamma^2 h c}{\lambda_u(1 + K^2/2)}.
\label{wundu}
\end{equation}
$\omega_{max}$ decreases with increase of the undulator parameter due to effective decrease of the electron velocity along $z$-axis caused by increase of the curvature of its trajectory in a strong field.
Note that $\omega_{max}(K \ll 1)$ in (\ref{wundu}) and $\omega_{max}(\kappa \ll 1)$ in (\ref{wwmax}) are the same if $\lambda_u = \lambda_0/2$.
So are the RTS photon energy spectrum (\ref{dsdw}) and the scattered photon distribution for a weak undulator radiation (see ref.~\cite{Hoffman}).
The only difference between RTS and weak undulator radiation is the difference between head sea and standing waves.

If we replace $B_0(z)$ by $B_y(z)$ in (\ref{PLB0}) and use the result in Eq.~(\ref{WCW}), we obtain the average number of RTS events in a long ($L\gg\lambda_u$) undulator:
\begin{equation}
W = \frac{1}{2}\cdot\frac{\Thomson}{\mu_0 h c} B_u^2 \lambda_u L 
= \frac{2\pi\alpha}{3}\frac{L}{\lambda_u}K^2,
\label{Wund}
\end{equation} 
where  $L$ is an undulator length and $\alpha$ is the fine structure constant. 
Factor $\frac{1}{2}$ in Eq.~(\ref{Wund}) comes from the decrease of equivalent photon density in a standing wave.

Equations (\ref{wave}--\ref{tyave}) can be used for undulator, bearing in mind the substitutions $\lambda_0 = 2\lambda_u$ and $\omega_0 = hc/2\lambda_u$.
The average electron energy loss in an undulator is determined as a product of Eq.~(\ref{Wund}) and $\left<\omega\right>$ from Eq.~(\ref{wave}):
\begin{equation}
\frac{\left<\Delta E\right>_u}{E} = \frac{\alpha}{3}\gamma \left( \frac{B_u}{B_c}\right)^2 \frac{L}{\labar_c},
\label{energy_loss_in_undulator}
\end{equation}
where $E=\gamma m c^2$ and $\labar_c = \lambda_c/2\pi$.

\subsection{Emittance Excitation}

If an electron beam with the horizontal emittance $\epsilon_x$ passes through an undulator (or laser beam), located in the dispersion-free section of the beam orbit, beam emittance is increased by radiation.
For a single case of RTS the average increase of $\epsilon_x$ depends on the beam $\beta$-function~\cite{bruck1966accelerateurs}:
\begin{equation}
\left<\Delta\epsilon_1\right>_u = \frac{\beta}{2}  {\left<\theta_x^2\right>}
= \frac{3}{20}\beta\left(\frac{\lambda_c}{\lambda_u}\right)^2,
\label{TT1}
\end{equation}
where $\theta_x = \theta_\varepsilon\cos\varphi$ ($\theta_\varepsilon\cos\varphi$ was introduced in Eq.~(\ref{txave})).
If we locate the centre of an undulator at the local minimum of $\beta$-function, $\beta_0$, we have $\beta(z)=\beta_0+z^2/\beta_0$.
Consequently, the average value of $\beta$-function inside this undulator is $\left<\beta\right>=\beta_0\left[1+(L/\beta_0)^2/12\right]$.
One can write the expression for the emittance excitation in an undulator as the product of Eq.~(\ref{Wund}) and Eq.~(\ref{TT1}):
\begin{equation}
\left<\Delta\epsilon_x\right>_u
=  \frac{\pi\alpha}{10} \left( \frac{B_u}{B_c}\right)^2  \frac{L}{\lambda_u} \left<\beta\right>.
\label{emittance_growth_in_undulator}
\end{equation}
Note that the vertical emittance excitation is smaller, $\left<\Delta\epsilon_y\right> = 2/3\left<\Delta\epsilon_x\right>$ according to Eq.~(\ref{tyave}).
In the longitudinal dimension, the energy spread $\sigma_E$ in the electron beam is increased by RTS.
This effect could be evaluated as the product of Eq.~(\ref{Wund}) and Eq.~(\ref{w2ave}): 
\begin{equation}
\frac{\left< \Delta \sigma_E^2 \right>_u}{E^2} = \frac{14\pi\alpha}{15} \gamma^2 \left( \frac{B_u}{B_c}\right)^2 \frac{L}{\lambda_u}.
\label{energy_spread_growth_in_undulator}
\end{equation}

\section{Wiggler}

Let us now switch to the consideration of a wiggler, i.~e. an undulator with $K\gg1$.
The standard approach for this case was taken from~\cite{Chao:384825}.
The following expressions describe a wiggler field $B_y(z)$ and local radius of an electron trajectory $\rho(z)$:
\begin{equation}
\begin{split}
& B_y(z) =  B_w \sin(z/\labar_w); \hspace{1em} \frac{1}{\rho(z)}  =  \frac{\sin(z/\labar_w)}{\rho_0}, \text{ where} \\
& \hspace{1em} \labar_w = \frac{\lambda_w}{2\pi}, \hspace{2em}   \rho_0 = \frac{E}{e c B_w} = \frac{(B\rho)}{B_w}
\end{split}
\label{wiggler}
\end{equation}
and $(B\rho)=P/e=\gamma \labar_c B_c$ is the electron rigidity.
Average electron energy loss in a wiggler is determined by its radiation integral $I_2^w$~[m$^{-1}$] as follows:
\begin{equation}
\frac{\left<\Delta E\right>_w}{E} =  \frac{2}{3}\alpha\labar_c\gamma^3 I_2^w, \;\;\text{ where }
I_2^w = \int\!\!\frac{dz}{\rho^2(z)}.
\label{I2}
\end{equation}
Integration along the wiggler of length $L\gg\lambda_w$ yields the same result as was obtained in Eq.~(\ref{energy_loss_in_undulator}) for RTS:
\begin{equation}
\frac{\left<\Delta E\right>_w}{E} = \frac{\alpha}{3}\gamma \left( \frac{B_w}{B_c}\right)^2 \frac{L}{\labar_c}.
\label{energy_loss_in_wiggler}
\end{equation}

A spontaneous emission of the synchrotron radiation photons leads to the excitation of the betatron oscillations, while this process takes place in the orbit regions with the non-zero dispersion functions, $D$ and $D'$.
If a wiggler is located in the dispersion-free section of the orbit, it will induce the dispersion
\begin{equation}
D_x(z)  = \frac{\labar_w^2}{\rho_0} \sin(z/\labar_w), \;\;\;\; 
D'_x(z) = \frac{\labar_w}{\rho_0}\cos(z/\labar_w).
\label{wiggler_dispersion} 
\end{equation}
The excitation of horizontal emittance is determined by radiation integral $I_5^w$~[m$^{-1}$]:
\begin{equation}
\left<\Delta \epsilon_x\right>_w =  \frac{55\alpha}{48\sqrt{3}}\labar_c^2\gamma^5 I_5^w, \text{ where }
I_5^w = \int\!\frac{\mathcal{H}(z)dz}{|\rho^3(z)|}.
\label{I5}
\end{equation}
If the centre of a wiggler ($z=0$) is located in a local minimum of the $\beta$-function $\beta(0)=\beta_0$, $\mathcal{H}(z)$ is:
\begin{equation}
\mathcal{H}(z) = \left( \beta_0^2 D'^2 + (sD'-D)^2 \right)\big/\beta_0.
\label{H(s)}
\end{equation}
Integration yields the result:
\begin{equation}
\left<\Delta \epsilon_x\right>_w = \frac{11\alpha}{18\sqrt{3}} \left( \frac{B_w}{B_c}\right)^2 \frac{L}{\lambda_w} \left<\beta\right> K^3,
\label{emittance_growth_in_wiggler}
\end{equation}
where $\left<\beta\right>$ is the same as in Eq.~(\ref{emittance_growth_in_undulator}).
The increase in the energy spread of the electron beam is determined by the radiation integral $I_3^w$~[m$^{-2}$]:
\begin{equation}
\frac{\left< \Delta \sigma_E^2 \right>_w}{E^2} =  \frac{55\alpha}{48\sqrt{3}}\labar_c^2\gamma^5 I_3^w, \text{ where }
I_3^w = \int\!\!\frac{dz}{|\rho^3(z)|}.
\label{I3}
\end{equation}
After integration we obtain:
\begin{equation}
\frac{\left< \Delta \sigma_E^2 \right>_w}{E^2} = \frac{55\alpha}{18\sqrt{3}} \gamma^2 \left( \frac{B_w}{B_c}\right)^2 \frac{L}{\lambda_w} K .
\label{energy_spread_growth_in_wiggler}
\end{equation}

Comparing Eq.~(\ref{energy_loss_in_undulator}) and Eq.~(\ref{energy_loss_in_wiggler}) one can see that the average electron energy losses in either undulator or wiggler are the same and independent of the analysis approach.
But the results for emittance excitation are different due to essential differences in the RTS and the synchrotron radiation properties. 
The RTS-based expressions (\ref{emittance_growth_in_undulator}, \ref{energy_spread_growth_in_undulator}) are valid only for $K\ll1$, where the standard radiation integrals show zero emittance excitation rate ($\left<\Delta \epsilon_x\right>\propto K^3$, Eq.~(\ref{emittance_growth_in_wiggler})).
This is because RTS is not included to the radiation integrals and it should be if we want them to be universal. 
Since the transverse emittance and the mean squared energy deviation represent the squared amplitudes of betatron and synchrotron oscillations, the correct way of combining corresponding contributions from RTS and synchrotron radiation is their direct summation.
This yields the something like:
\begin{eqnarray}
\left<\Delta \epsilon_x \right>& \simeq &
\frac{11\alpha}{18\sqrt{3}}  \left( \frac{B_w}{B_c}\right)^2  \frac{L}{\lambda_w} \left<\beta\right> \left(0.89 +  K^3\right),\\
\frac{\left<\Delta \sigma_E^2 \right>}{E^2} & \simeq &
\frac{55\alpha}{18\sqrt{3}}\left( \frac{B_w}{B_c}\right)^2 \frac{L}{\lambda_w}\;\; \gamma^2 \left(1.66 + K\right).
\end{eqnarray}
These expressions can be used to update an undulator/wiggler radiation integrals:
\begin{eqnarray}
I_2^w & = & \frac{L}{\lambda_w}\left[\frac{B_w}{B\rho}\right]\frac{\pi K}{\gamma} = \frac{L}{2}\left[\frac{B_w}{B\rho}\right]^2, \label{I2W}\\
I_3^w & = & \frac{L}{\lambda_w}\left[\frac{B_w}{B\rho}\right]^2\frac{\left(1.66 + K\right)}{\gamma},  \label{I3W}\\
I_5^w & = & \frac{L}{\lambda_w}\left[\frac{B_w}{B\rho}\right]^2\frac{8}{15}\frac{\left<\beta\right>\left(0.89 +  K^3\right)}{\gamma^3}  \label{I5W}.
\end{eqnarray}

\section{Radiation Damping}

\begin{figure}[h!]
\centering
\includegraphics[width=\linewidth]{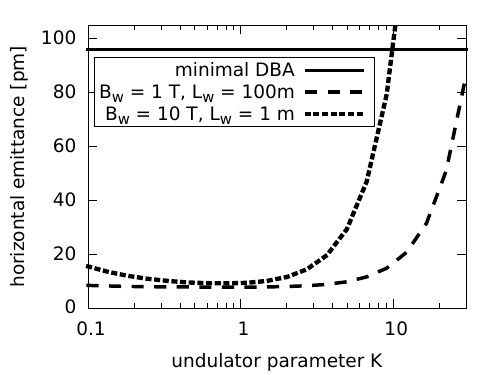}
\caption{Horizontal emittance $\epsilon_x$ vs $K$, $\left<\beta\right>=10$~m.}
\label{fig:xemittance}
\end{figure}

In an electron storage ring the equilibrium beam emittances are formed by radiation damping while the longitudinal momentum losses are replaced by rf-system~\cite{Chao:384825}:
\begin{eqnarray}
\left(\frac{\sigma_E}{E}\right)^2 & = & C_q \gamma^2 \frac{I_3}{ 2 I_2 + I_{4x}  + I_{4y}}, \\
\epsilon_{x,y}             & = & C_q \frac{\gamma^2}{J_{x,y}} \frac{I_{5x,y}}{I_2}\;\; \left(C_q = \frac{55\labar_c}{32\sqrt{3}}\right).
\end{eqnarray}
The concept of damping wigglers application is to increase the radiation integral $I_2$ and therefore decrease the equilibrium emittances.
Neglecting $I_4$ and assuming $J_x=1$ the following expressions describe this equilibrium for a storage ring with damping wigglers:
\begin{eqnarray}
\left(\frac{\sigma_E}{E}\right)^2 & = & \frac{C_q \gamma^2}{2} \frac{I_3 + I_3^w}{I_2 +I_2^w}, \label{sEE}\\
\epsilon_{x}                      & = & C_q \gamma^2 \frac{I_5 + I_5^w}{I_2 +I_2^w}, \label{Ex}
\end{eqnarray}
where $I_{2,3,5}$ are radiation integrals of the storage ring itself and $I^w_{2,3,5}$ are due to the damping wigglers.
The limiting case when $I_{i}^w \gg I_{i}\;(i =2,3,5)$ gives the emittance values in the wiggler-dominated storage ring:
\begin{eqnarray}
\left(\frac{\sigma_E}{E}\right)^2 & = & \frac{C_q}{\lambda_w} \gamma \left(1.66 + K\right), \label{WDE1}\\
\epsilon_{x}                      & = & \frac{C_q }{\lambda_w}\frac{16}{15}\frac{\left<\beta\right>}{\gamma}\left(0.89 +  K^3\right).\label{WDE2}
\end{eqnarray}
For $K=0$ case Eqs.~(\ref{WDE1}, \ref{WDE2}) yield the same results as were obtained in \cite{PhysRevLett.80.976}.
However the value of $I_5^w$ if $(K=0)$ is so small that practically it can not be comparable to $I_5$ of a ring 
(this issue seems have not been properly taken into account in \cite{PhysRevLett.80.976}).
Let us consider an imaginary storage ring with the DBA lattice~\cite{Chao:384825} and the parameters:
\begin{itemize}
\item Beam energy -- 2~GeV ($\gamma=4\cdot10^3$).
\item 100 bending dipoles with $B=0.1$~T.
\item Minimal DBA emittance $\epsilon_x=96$~pm.
\end{itemize}
In Fig.~\ref{fig:xemittance} the dashed curve represents the horizontal emittance change due to wiggler damping section of length $L_w=100$~m and $B_w=1$~T.
One can see that the damping effect does not depend on $K$ until it reaches the value of $K\simeq10$, where the impact of the constant factor 0.89 in Eq.~(\ref{I5W}) is negligible.
The situation when this impact could become noticeable is represented by the dotted curve in Fig.~\ref{fig:xemittance}. 
The parameters of this curve are taken as $L_w=1$~m and $B_w=10$~T in order to keep the same value of the damping integral $I_2^w$.
This dotted curve shows small decrease in equilibrium emittance starting from $K=0.1$ up to $K\simeq1$.
The reason is that in this range $I_5^w$ is inverse proportional to the wiggler period $\lambda_w$, see Eq.~(\ref{I5W}). 
The parameter combination [$K=1$, $B_w=10$~T] means that $\lambda_w\simeq1$~mm and so it is beyond the existing technologies.
According to Eq.~(\ref{PLB0}) this is the case when $\sim10^{11}$~W of radiation power is focused to 1~mm transverse size along 1~m length.
Though only 1~m of a damping section length provides the same damping effect as an 100~m section of regular wigglers.
\begin{figure}[h!]
\includegraphics[width=\linewidth]{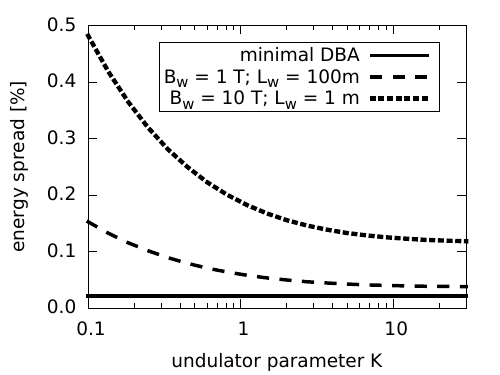}
\caption{Beam energy spread $\sigma_E/E$ vs $K$.}
\label{fig:eemittance}
\end{figure}

In the longitudinal dimension the equilibrium beam energy spread is shown in Fig.~\ref{fig:eemittance} for the same parameters as they were in Fig.~\ref{fig:xemittance}.
The energy spread is always growing with the decrease in the undulator parameter $K$.
It happens due to the valuable impact of the constant factor 1.66 in Eq.~(\ref{I3W}).

\section{Conclusion}
We have examined the impact of relativistic Thomson scattering to the radiation integrals of undulators and wigglers.
The solution is suggested for the correct evaluation of these integrals for arbitrary values of undulator parameter.
When $K=0$, these results are in agreement with the ones obtained earlier \cite{PhysRevLett.80.976,Lebedev:2004yz}.
However, we used a somewhat different approach, provide an opportunity for optimisation of the damping parameters and also a proper account of the influence of the direction of wave polarisation on the excitation of betatron oscillations, see Eqs.~(\ref{txave}) and (\ref{tyave}).
We point out the possibility of a substantial reduction of the length of the cooling section, preserving its effectiveness by increasing the field and reducing the period.
However, the parameter combination like $B_w\simeq10$~T and $\lambda_w\simeq1$~mm is presently beyond the existing technology limits for either laser or undulator cases.
But if this limit could be overcome, the pending amendment will have a decisive role.

\begin{acknowledgments}
The author wants to express his gratitude to Eugene Levichev for inspiration of these studies and for useful discussions.
This work was supported by Russian Science Foundation (project N 14-50-00080).
\end{acknowledgments}

\bibliography{undu}

\end{document}